\documentclass[reprint,superscriptaddress,amsmath,amssymb,aps,prl]{revtex4-2}

\usepackage{graphicx}
\usepackage{dcolumn}
\usepackage{bm}
\usepackage{hyperref}
\usepackage{color}

\newcommand{\dd}{\mathrm{d}}

\begin{document}

\title{Holography for stress-energy tensor flows}

\affiliation{Center for Theoretical Physics and College of Physics, Jilin University, 
Changchun 130012, China}

\author{Xi-Yang Ran}
\email{ranxy23@mails.jlu.edu.cn}
\affiliation{Center for Theoretical Physics and College of Physics, Jilin University, 
Changchun 130012, China}

\author{Feng Hao}
\email{haofeng22@mails.jlu.edu.cn}
\affiliation{Center for Theoretical Physics and College of Physics, Jilin University, 
Changchun 130012, China}

\author{Hao Ouyang}
\email{haoouyang@jlu.edu.cn}
\affiliation{Center for Theoretical Physics and College of Physics, Jilin University, Changchun 130012, China}

\begin{abstract}
We study the holographic description for general stress-energy tensor deformations in arbitrary dimensions using the metric flow approach. Mixed boundary conditions corresponding to these deformations emerge from solutions to the metric flow equations. To test this proposal, we analyze planar anti-de Sitter black holes with such boundary conditions and find that the deformed energies satisfy flow equations consistent with the field theory interpretation. We further derive the commuting condition for stress-energy tensor deformations and extend the mixed boundary condition description to accommodate families of commuting deformations.
\end{abstract}

\maketitle


\section{Introduction}
Extending the AdS/CFT correspondence \cite{Maldacena:1997re,Gubser:1998bc,Witten:1998qj} via controlled deformations can offer new insights into quantum gravity and holography.
One notable example is the two-dimensional $T\overline{T}$ deformation \cite{Zamolodchikov:2004ce,Cavaglia:2016oda,Smirnov:2016lqw}, which has attracted considerable attention due to its remarkable properties.  
For comprehensive reviews, see Refs. \cite{Jiang:2019epa,He:2025ppz}.
It was proposed in Ref. \cite{McGough:2016lol} that turning on the $T\overline{T}$ deformation is holographically dual to introducing a finite radial cutoff surface in the bulk. 
However, this cutoff prescription is only applicable to one sign of the deformation parameter.
For the opposite sign, a glue-on proposal was introduced in Ref. \cite{Apolo:2023vnm}. 
Both the cutoff and glue-on approaches are restricted to the pure gravity sector.
A more flexible proposal arises from imposing mixed boundary conditions at infinity \cite{Guica:2019nzm}, which works for both signs of the deformation parameter and accommodates matter fields in the bulk.
Beyond these proposals, the random geometry formulation of $T\overline{T}$ \cite{Cardy:2018sdv} has inspired an alternative holographic interpretation \cite{Hirano:2020nwq}.

Considering general dimensions, the $T\overline{T}$ deformation can be generalized to stress-energy tensor deformations driven by composite operators built from the stress-energy tensor \cite{Taylor:2018xcy, Bonelli:2018kik, Conti:2018jho, Cardy:2018sdv, Ferko:2019oyv, Babaei-Aghbolagh:2020kjg, Conti:2022egv, Hou:2022csf, Ferko:2023sps,Ferko:2023wyi, Ferko:2024zth,Babaei-Aghbolagh:2024hti,Tsolakidis:2024wut}.
These deformations retain some of the geometric features of the $T\overline{T}$ deformation \cite{Cardy:2018sdv, Conti:2018tca}, and their effects can be reformulated as an auxiliary flow in the background metric \cite{Conti:2022egv,Morone:2024ffm, Ran:2024vgl}.
This metric flow picture provides a powerful tool for establishing connections between quadratic stress-energy tensor deformations and Ricci flows \cite{Brizio:2024arr,Morone:2024sdg}.

In this Letter, we intend to explore the holographic description for general stress-energy tensor deformations in arbitrary dimensions.
While the cutoff prescription has been generalized to higher-dimensional $T\overline{T}$-like deformations\cite{Taylor:2018xcy,Hartman:2018tkw,Shyam:2018sro,Caputa:2019pam,Belin:2020oib,Parvizi:2025shq, Parvizi:2025wsg}, we instead employ the mixed boundary condition approach, applicable to more general deformations.
Motivated by a key observation that the mixed boundary condition for the $T\overline{T}$ deformation is closely related to the metric flow approach in Ref. \cite{Conti:2022egv}, we propose that this connection extends to general stress-energy tensor deformations.
The metric flow equations provide a systematic method for identifying the corresponding mixed boundary conditions.
As a concrete example, we analyze planar anti-de Sitter (AdS) black holes. We find that the resulting spectrum of deformed energies satisfies flow equations consistent with those derived from large-$N$ deformed field theories.

As a related topic, we explore the commutativity of stress-energy tensor deformations.
The commutativity between the $T\overline{T}$ and root-$T\overline{T}$ \cite{Rodriguez:2021tcz, Babaei-Aghbolagh:2022uij,Ferko:2022iru,Conti:2022egv,Ferko:2022cix,Babaei-Aghbolagh:2022leo} deformations has been used in Ref. \cite{Ebert:2023tih} to determine the holographic mixed boundary condition for the root-$T\overline{T}$ deformation. In this work, we derive a general commuting condition for stress-energy tensor deformations using the metric flow approach.
Since both the mixed boundary conditions and commuting condition can be derived from the metric flow equations, commuting deformations naturally lead to compatible mixed boundary conditions, opening new possibilities to construct richer and more flexible holographic models.

\section{Metric flows}

We begin by reviewing the metric flow prescription for stress-energy tensor deformations and clarifying its relation to the variational principle method for the mixed boundary conditions. 
The action of a $d$-dimensional field theory deformed by an operator constructed from the stress-energy tensor obeys the differential equation 
\begin{equation}
    \frac{\partial S_\tau(g_{\mu\nu},\phi)}{\partial\tau}=\int\dd^dx\sqrt{g}O,
    ~~~g=|\det g_{\mu\nu}|, 
\end{equation}
where $\phi$ denotes the collection of matter fields, $\tau$ is the deformation parameter, and the deformation operator $O$ is a Lorentz-invariant function of the stress-energy tensor $T^\mu_{~\nu}=T^{\mu\sigma}g_{\sigma\nu}$ defined by 
\begin{align}\label{eq:def of stress tensor}
    T^{\mu\nu}=\frac{-2}{\sqrt{g}}\frac{\delta S_\tau(g_{\mu\nu},\phi)}{\delta g_{\mu\nu}}.
\end{align}
The operator $O$ may also have explicit dependence on $\tau$ or other dimensionful constants, as in $T\overline{T}+\Lambda_2$ \cite{Lewkowycz:2019xse,Gorbenko:2018oov} and in the deformations studied in Refs. \cite{Tsolakidis:2024wut,Blair:2024aqz}. 
We call a deformation stationary if $O$ has no explicit dependence on $\tau$, and homogeneous if $O$ is a homogeneous 
function of $T^\mu_{~\nu}$ so that 
\begin{equation}
    T^\mu_{~\nu}\frac{\partial O}{\partial T^\mu_{~\nu}}=m O,
\end{equation}
where $m$ is the degree of $O$.

As noted in Refs. \cite{Conti:2022egv, Morone:2024ffm, Ran:2024vgl}, a stress-energy tensor flow can alternatively be interpreted as an auxiliary flow in the space of the background metrics
\begin{align}
    \frac{\dd g_{\mu\nu}}{\dd\tau}=&2\frac{\partial O}{\partial T^{\mu\nu}},\label{flow-g-1}\\
    \frac{\dd T^{\mu\nu}}{\dd\tau}=&(T^{\rho\sigma}g^{\mu\nu}-T^{\mu\nu}g^{\rho\sigma})\frac{\partial O}{\partial T^{\rho\sigma}}-2\frac{\partial O}{\partial g_{\mu\nu}}-Og^{\mu\nu},\label{flow-T-1}
\end{align}
arising from a classical dynamical equivalence between the deformed theory and the undeformed theory with a modified background metric. 
Consequently, the deformed action for a stationary homogeneous deformation can be obtained as 
\begin{equation}\label{defaction}
 S_\tau(g_{\mu\nu},\phi)
 =S_0(g_{\mu\nu,0},\phi)
 +(1-m)\tau \int\dd^dx\sqrt{g_0}O(T^\mu_{~\nu,0}),
\end{equation}
where $g_{\mu\nu,0}$ and $T^\mu_{~\nu,0}$ are, respectively, the initial values of $g_{\mu\nu}$ and $T^\mu_{~\nu}$ at $\tau=0$. 
A derivation of \eqref{defaction}, along with its generalization to broader classes of deformations, is given in Ref. \cite{Morone:2024sdg}. 
Alternatively, the flow equations  \eqref{flow-g-1} and \eqref{flow-T-1} can be understood as the flows along the characteristic curves in the procedure of solving the classical deformed Lagrangian with the method of characteristics \cite{Hou:2022csf}.

Within the AdS/CFT correspondence, multitrace deformations of the boundary theory correspond to modified boundary conditions for the dual bulk fields in AdS \cite{Klebanov:1999tb,Witten:2001ua,Berkooz:2002ug,Mueck:2002gm,Gubser:2002zh,Hartman:2006dy,Diaz:2007an,Papadimitriou:2007sj}. 
Boundary conditions for the $J\bar{T}$ and $T\bar{T}$ deformations were analyzed via the variational principle in Refs. \cite{Bzowski:2018pcy} and \cite{Guica:2019nzm}, respectively. 
This approach was later applied to more general stress tensor deformations in AdS$_3$/CFT$_2$ in Ref. \cite{Ebert:2023tih}.
Here, we relate it to the flow equations \eqref{flow-g-1} and \eqref{flow-T-1}. 
We consider a large-$N$ field theory deformed by a multitrace operator $f(\mathcal{O})$ where $\mathcal{O}$ is a single-trace operator dual to a fundamental bulk field. 
In the large-$N$ limit, the deformed generating functional $W_\tau[J]$ is related to the undeformed one as
\begin{equation}
    W_\tau[J_\tau]=W_0[J_0]+\tau \int \dd^d x (f-\sigma f'(\sigma)), 
\end{equation}
where $J_\tau$ and $J_0$ are the sources coupled to $\mathcal{O}$ in the deformed and undeformed theories, respectively, related by $J_\tau=J_0-\tau f'(\sigma)$. 
Here, $\sigma=\delta W_0/\delta J_0$ is the undeformed expectation value of $\mathcal{O}$. 
The details of derivation can be found in Ref. \cite{Papadimitriou:2007sj} (see also Ref. \cite{Ebert:2023tih}). 
Taking the limit $\tau\rightarrow 0$, we get the differential equations
\begin{align}
    \partial_\tau W_\tau[J_\tau]&= \int \dd^d x (f-\sigma f'(\sigma)), \label{diffW}\\
    \partial_\tau J_\tau&=- f'(\sigma).\label{diffJ}
\end{align}
In stress-energy tensor deformations, the source $J_\tau$ coupled to the deformed stress tensor $T^{\mu\nu}$ is the deformed metric $g_{\mu\nu}$. 
As discussed in Ref. \cite{Ebert:2023tih}, the function $f$ depends on both single-trace operators and their sources, leading to more complicated behavior. 
Therefore, it is more convenient to use the varied version of \eqref{diffW} for a stress tensor deformation $O(T^\mu_{~\nu})$ 
\begin{equation}
    \begin{split}
        &\frac{1}{2}\frac{\partial}{\partial\tau}\left( \int \dd^d x \sqrt{g} T_{\mu\nu} \delta g^{\mu\nu}\right)\\
        =
        & \delta \left(  \int \dd^d x \sqrt{g} ( O-T^\mu_{~\nu}\frac{\partial O}{\partial T^\mu_{~\nu}}) \right)  .
    \end{split}
\end{equation}
With a slight abuse of notation, here we use the same symbols for the operators $T^\mu_{~\nu}$ and their expectation values $\langle T^\mu_{~\nu}\rangle$, justified by large-$N$ factorization.
By matching varied terms and their corresponding coefficients, this yields 
\begin{align}
    \frac{1}{2}\partial_\tau(\sqrt{g} T^{\mu}_{~\nu}) g^{\nu\sigma}\delta g_{\sigma\mu}=& (T^\mu_{~\nu}\frac{\partial O}{\partial T^\mu_{~\nu}}- O)\delta\sqrt{g}, \\
    \frac{1}{2}\sqrt{g} T^{\mu}_{~\nu} \partial_\tau(g^{\nu\sigma}\delta g_{\sigma\mu})=& \sqrt{g}\delta(T^\mu_{~\nu}\frac{\partial O}{\partial T^\mu_{~\nu}}- O), 
\end{align}
from which one can derive the same flow equations \eqref{flow-g-1} and \eqref{flow-T-1} obtained from the classical analysis. 
This is expected, since the computation of the generating functional reduces to a classical saddle point in the large-$N$ limit. 
As a nontrivial check, the solution of the flow equations \eqref{flow-g-1} and \eqref{flow-T-1} for the root-$T\bar T$ operator gives the mixed boundary condition proposed in Ref. \cite{Ebert:2023tih} (see also Refs. \cite{Ebert:2024fpc,Tian:2024vln,Babaei-Aghbolagh:2024hti}).

\section{Solution to the flow equations}

Equations \eqref{flow-g-1} and \eqref{flow-T-1} can be solved perturbatively in small $\tau$, with initial values given by $g_{\mu\nu,0}$ and $T^{\mu}_{~\nu,0}$. 
Both $g^{\mu\alpha,0}g_{\alpha\nu}$ and $T^\mu_{~\nu}$ admit formal power series expansions in $\tau$, and the coefficients are functions of $T^{\mu}_{~\nu,0}$. 
As a result, $g^{\mu\sigma,0}g_{\sigma\nu}$, $T^{\mu}_{~\nu}$ and $T^\mu_{~\nu,0}$ can be diagonalized simultaneously by a $\tau$-independent matrix. 
We denote the eigenvalues of $g^{\mu\sigma,0}g_{\sigma\nu}$ and $T^\mu_{~\nu}$ as $\omega_\alpha$ and $t_\alpha$, respectively. 
In terms of these eigenvalues, $O$ can be treated as a function of $t_\alpha$, and the flow equations take the form
\begin{align}
   \frac{\dd \omega_\alpha}{\dd\tau}=&2\frac{\partial O}{\partial t_\alpha}\omega_\alpha ,\label{floweqmetric}
   \\
   \frac{\dd t_\alpha}{\dd\tau}=&-O
   +\sum_{\beta} t_\beta \frac{\partial O}{\partial t_\beta}
   -t_\alpha \sum_{\beta} \frac{\partial O}{\partial t_\beta}.
   \label{floweqt}
\end{align}
Throughout the paper, repeated indices of eigenvalues are not implicitly summed over, while repeated indices of tensors are implicitly summed over their respective ranges.

For a stationary homogeneous deformation, the solution to the flow equation for $t_\alpha$ (\ref{floweqt}) with initial condition $t_{\alpha,0}$ is given by
\begin{equation}\label{eq:solution for t_a}
    t_\alpha=\frac{t_{\alpha,0}+s}{\sqrt{g/g_0}} ,\quad \sqrt{g/g_0}=\left(\frac{O|_{t_\alpha\rightarrow t_{\alpha,0}+s}}{O|_{t_\alpha\rightarrow t_{\alpha,0}}}\right)^{\frac{1}{m-1}},
\end{equation}
where $s=\tau(m-1) O|_{t_\alpha\rightarrow t_{\alpha,0}}$. 
For marginal deformations, the solution can be obtained by taking the limit $m\rightarrow 1$. 
This solution can be obtained by following a similar strategy used in Ref. \cite{Hou:2022csf}. 
First, one can show that the quantity $\sqrt{g} O(t_\alpha)$ is invariant along the flow by using the flow equations \cite{Morone:2024sdg}. 
Then, one finds that
\begin{equation}
    \frac{\dd}{\dd\tau}  \left( \sqrt{g}t_\alpha\right)
    =
    \sqrt{g}\big(\sum_{\beta} t_\beta\partial_{t_\beta}-1\big)O
    =
    (m-1)\sqrt{g}O
\end{equation}
remains invariant along the flow, allowing $\sqrt{g}t_\alpha$ to be determined. 
Finally, $\sqrt{g}$  can be determined from the relation $\sqrt{g} O(t_\alpha)=\sqrt{g_0} O(t_{\alpha,0}) $ and the homogeneity of $O$. 
Once $t_\alpha$ are determined, we can integrate equations \eqref{floweqmetric} to get $\omega_\alpha$.

While explicit solutions for general deformations are not available in closed form, the formula 
\begin{equation}\label{difft}
    \frac{\dd}{\dd\tau}  \left( \sqrt{g}(t_\alpha-t_\beta)\right)  =0,
\end{equation}
for any two $t$-eigenvalues, will be useful for our analysis.

\section{Holography}

We now turn to mixed boundary conditions associated with the stress-energy tensor in holography. 
The metric of an asymptotically AdS spacetime can be written in the Graham-Fefferman coordinate system 
\begin{equation}
    \dd s^2 =\frac{\ell^2 \dd\rho^2}{4\rho^2}+\frac{1}{\rho} \gamma_{\mu\nu}(\rho) \dd x^\mu \dd x^\nu,
\end{equation}
where
\begin{equation}\label{metricexpand}
    \gamma_{\mu\nu}(\rho)= \gamma_{\mu\nu}^{(0)}+\cdots+\gamma_{\mu\nu}^{(d)} \rho^{d/2}  +\rho^{d/2}h_{\mu\nu}\log \rho +\cdots~.
\end{equation}
From now on, we use $\mu,\nu,\cdots=0,\cdots,d-1$ to label spacetime coordinates and $i,j,\cdots=1,\cdots,d-1$ to label spatial coordinates. 
According to the AdS/CFT dictionary \cite{Balasubramanian:1999re,deHaro:2000vlm}, the boundary metric and stress-energy tensor in the undeformed field theory are related to the coefficients $\gamma_{\mu\nu}^{(n)}$ as
\begin{equation}\label{holo-und}
    g_{\mu\nu,0}=\gamma_{\mu\nu}^{(0)},\quad T_{\mu\nu,0}=\frac{d \ell^{d-3}}{16 \pi G } \gamma_{\mu\nu}^{(d)} +X_{\mu\nu}(\gamma_{\mu\nu}^{(n)}),
\end{equation}
where $X_{\mu\nu}$ is a function of $\gamma_{\mu\nu}^{(n)}$ with $n<d$ encoding the conformal anomalies of the boundary conformal field theory. 
After turning on the stress-energy deformation, the boundary metric and stress-energy tensor can be obtained by solving the flow equations with \eqref{holo-und} as the initial values. 
In this way, the solution to the flow equation provides a holographic prescription for the mixed boundary condition of the stress-energy tensor.

The mixed boundary condition proposals of $T \bar T$ and root-$T\bar T$ deformation were verified respectively in Refs. \cite{Guica:2019nzm} and \cite{Ebert:2023tih} by matching the energy of a deformed Ba\~nados-Teitelboim-Zanelli (BTZ) black hole with the field theory answer. 
We now verify the mixed boundary condition proposal for general stress-energy tensor deformations in planar AdS black holes, following similar steps as in Refs. \cite{Guica:2019nzm,Ebert:2023tih}. 
The metric of a planar AdS black hole in $d + 1$ dimensions is
\begin{equation}
    \begin{split}
        \dd s^2 = &\frac{\ell^2}{z^2} \bigg[- \left(1 - \frac{z^d}{z_0^d}\right) \dd t^2 + \left(1 - \frac{z^d}{z_0^d}\right)^{-1} \dd z^2 
        \\
        &+ \sum_{i=1}^{d-1} \dd x^i \dd x^i \bigg].    
    \end{split}
\end{equation}
The transformation $z= \ell \sqrt{\rho }  (\mathcal{L} \rho ^{d/2}+1 )^{-2/d}$ with $z_0=4^{-1/d} \ell\mathcal{L}^{-1/d}$ brings the metric to the Fefferman-Graham form 
\begin{equation}
    \begin{split}
        \dd s^2 =& \frac{\ell^2 \dd \rho^2}{4\rho^2}-\frac{ \left(\mathcal{L} \rho ^{d/2}-1\right)^2 \left(\mathcal{L} \rho ^{d/2}+1\right)^{\frac{4}{d}-2}}{\rho } \dd t^2
        \\
        &+\frac{ \left(\mathcal{L}\rho ^{d/2}+1\right)^{4/d}}{\rho } \sum_{i=1}^{d-1} \dd x^i \dd x^i.
    \end{split}
\end{equation}
Expanding the metric near the boundary $\rho\rightarrow 0$ and comparing with the expansion \eqref{metricexpand}, we find that
\begin{equation}
    \begin{split}
        \gamma_{\mu\nu}^{(0)}=&\mathrm{diag}(-1,1,...,1) ,\quad \gamma_{\mu\nu}^{(n)}=0,~0<n<d,
        \\
        \gamma_{\mu\nu}^{(d)}=&\frac{4\mathcal{L}}{d}\mathrm{diag}(d-1,1,...,1) . 
    \end{split}
\end{equation}
From the AdS/CFT dictionary, the undeformed boundary metric and stress-energy tensor can be read off as 
\begin{align}
    g_{\mu\nu,0}&=\gamma_{\mu\nu}^{(0)},
    \\
    T_{\mu\nu,0}&=\frac{d \ell^{d-3}}{16 \pi G } \gamma_{\mu\nu}^{(d)}=\mathrm{diag}((d-1) h,h,...,h),
\end{align}
where we introduce $h= {\ell^{d-3}}(4\pi G)^{-1}\mathcal{L}$ to simplify the expression. 
In this case, the boundary metric is flat and there are no gravitational conformal anomalies, so $X_{\mu\nu}=0$. 
The deformed boundary metric and stress tensor can be obtained as the solution to the flow equations
\begin{align}
    g^{\mu\alpha, 0}g_{\alpha\nu}=& \mathrm{diag}(\omega_0,\cdots,\omega_{d-1}),
    \\
    T^{\mu}_{~~\nu}=&\mathrm{diag}(t_0,\cdots,t_{d-1}).
\end{align}
It follows from \eqref{difft} that the spatial $t$-eigenvalues are equal, 
so we denote $t_i=t_x$ for $i=1,...,d-1$. 
For a stationary homogeneous deformation, we have
\begin{align}
    t_0=&\frac{1}{\sqrt{g/g_0}} \left((1-d)h+\tau(m-1)O(t_{\alpha,0})\right),
    \\
    t_x=&\frac{1}{\sqrt{g/g_0}}\left(h+\tau(m-1)O(t_{\alpha,0})\right). 
\end{align}

To obtain the spectrum, we need to work with coordinates such that the boundary metric takes the standard form: 
\begin{equation}
    g_{\mu\nu}\dd x^\mu \dd x^\nu=-\dd T^2+\sum_{i=1}^{d-1} \dd\phi^i \dd\phi^i. 
\end{equation}
This can be achieved by $t \rightarrow T /\sqrt{\omega_0}$ and 
$\phi_i \rightarrow x_i /\sqrt{\omega_i}$. 
To introduce nontrivial angular momenta, one can further make a Lorentz boost. Then the full coordinate transformation is
\begin{align}
    t=&\frac{1}{\sqrt{\omega_0}} \left( \Xi T+\sum_{i=1}^{d-1}a_i \phi^i \right),\label{eq:coordinate transformation 1}
    \\
    x^i=&\frac{1}{\sqrt{\omega_i}} \left( a_i T+\phi^i+\sum_{j=1}^{d-1}\frac{ a_i a_j}{\Xi+1}\phi^j \right) , \label{eq:coordinate transformation 2}
\end{align}
where $\Xi=(1+\sum_{i=1}^{d-1}a_i^2)^{1/2}$. 
To investigate off-diagonal components of the stress tensor, 
we compactify the $\phi$-coordinates on a nonorthogonal torus defined by identifying
\begin{equation}
    \phi^i\sim \phi^i+R^i_{~a},\quad i,a=1,\cdots,d-1. 
\end{equation}
The volume of the torus is $V=\det(R^i_{~a})$. 
The components of the stress-energy tensor in the new coordinate are
\begin{equation}
    \begin{split}
        T^0_{~0}=&\Xi^2 t_0-(\Xi^2-1)t_x,\quad  T^0_{~i}=a_i\Xi  (t_0-t_x),
        \\ 
        T^i_{~j}=&a_i a_j (t_x-t_0)+\delta^i_j t_x.  
    \end{split}
\end{equation}
The deformed energy and angular momentum are 
\begin{equation}
    E= -VT^0_{~0},\quad J_i= -V T^0_{~i}.
\end{equation}
To obtain the deformed energy for a given deformation, one should fix $\mathcal{L}$ (or $h$) and $a_i$ by identifying quantities that are not deformed. 
One such quantity is the area of the event horizon at $\rho=\mathcal{L}^{-d/2}$, 
\begin{equation}
    A=4^{\frac{d-1}{d}} V \mathcal{L}^{\frac{d-1}{d}} \Xi\prod_{i=1}^{d-1} {\omega_i^{-1/2}}.
\end{equation}
The event horizon area corresponds to the degeneracy of states, which should be unchanged under variation of $\tau$ and $R^i_{~a}$. 
The bulk angular momenta correspond to the momenta of the states in the field theory. 
The periodicity condition of the wave function requires that $J_i R^i_{~a}$ should be integer multiples of $2\pi$ and thus invariant under continuous change of $\tau$ and $R^i_{~a}$. 
Now, we have
\begin{align}
    V \mathcal{L}^{\frac{d-1}{d}} \Xi\prod_{i=1}^{d-1} {\omega_i^{-1/2}}=&\mathrm{constant},
    \\   
    V \Xi  (t_0-t_x) a_i R^i_{~a}=&\mathrm{constant}.
\end{align}
The constants are fixed by the undeformed energy and momenta. 
For a specific deformation, one can obtain the deformed energy by using the explicit expressions of $t_\alpha$ and $\omega_\alpha$ to solve the constraints. 
For generic deformations, we instead derive flow equations for the deformed energy. 
By taking derivatives of these constraints with respect to $\tau$ and $R^i_{~a}$, one can express the derivatives of $\mathcal{L}$ and $a_i$ as functions of $\mathcal{L}$ and $a_i$. 
We show in Supplemental Material \cite{supp} that
\begin{align}
    \partial_\tau  E &= V O(t_\alpha) \label{holo-E-flow},
    \\
    T^i_{~j}&=-\frac{1}{V} R^i_{~a}\frac{\partial E }{\partial R^j_{~a}}.\label{Tij}
\end{align}
Equation \eqref{Tij} is consistent with the field theory interpretation of the stress-energy tensor.

In a large-$N$ field theory, we assume that the expectation value of $O$ factorizes as $\langle O( T^\mu_{~\nu})\rangle=O(\langle T^\mu_{~\nu}\rangle)$. 
Then, the energy of an eigenstate of the energy and momentum operators satisfies the flow equation
\begin{equation}
    \partial_\tau  E = V O(\langle T^\mu_{~\nu}\rangle) .
\end{equation}
Therefore, \eqref{holo-E-flow} agrees with the flow equation of the energy obtained from the field theory in the large-$N$ limit.

\section{Commuting flows}

Now we investigate the holographic mixed boundary condition for commuting stress-energy tensor deformations. 
We need to derive the condition that two deformations commute with each other.
We consider two deformations driven by $O_1$ and $O_2$. 
When two deformations commute with each other, we have
\begin{equation}
   \partial_{\tau_1} \partial_{\tau_2}\omega_\alpha =
   \partial_{\tau_2} \partial_{\tau_1}\omega_\alpha,\quad 
   \partial_{\tau_1} \partial_{\tau_2}t_\alpha =
   \partial_{\tau_2} \partial_{\tau_1}t_\alpha.
\end{equation}
Using the flow equations, we show in the Supplemental Material \cite{supp} that the commuting condition is equivalent to
\begin{equation}\label{commuteeq}
    \begin{split}
        &C(O_1,O_2)\equiv \sum_{\alpha} (O_1 \frac{\partial O_2}{\partial t_\alpha}- O_2 \frac{\partial O_1}{\partial t_\alpha})
        \\&  -\sum_{\alpha,\beta}  (t_\beta-t_\alpha)
        \frac{\partial O_1}{\partial t_\beta} \frac{\partial O_2}{\partial t_\alpha}+\tilde\partial_{\tau_2}O_1-\tilde\partial_{\tau_1}O_2=0.  
    \end{split}
\end{equation}
Here, the symbol $\tilde\partial_{\tau_p} O_q$ refers only to the explicit dependence of $O_q$ on $\tau_p$. 
To simplify the analysis and uncover more interesting structures, we now restrict our discussion to stationary deformations. 
We observe that
\begin{itemize}
    \item[(i)] A deformation is called traceless if $\sum_\alpha \partial_{t_\alpha}O=0$. Traceless deformations mutually commute.
    \item[(ii)] Marginal deformations mutually commute.
    \item[(iii)] Traceless marginal deformations commute with arbitrary deformations, which is consistent with results in Ref. \cite{Babaei-Aghbolagh:2024hti}.
\end{itemize}
For a given deformation with a generic operator $O_1$, one can obtain commuting deformations by solving $C(O_1,O_2)=0$, which is a linear partial differential equation for $O_2$. 
Using the method of characteristics, a general solution can be written as
\begin{equation}\label{O2-O1G}
    \begin{split}
        &O_2(t_\alpha)\\
        =&O_1(t_\alpha) G \left(\frac{t_1-t_2}{O_1(t_\alpha)},\frac{t_2-t_3}{O_1(t_\alpha)},\cdots,\frac{t_{d-1}-t_d}{O_1(t_\alpha)}\right), 
    \end{split}
\end{equation}
where $G$ is an arbitrary function.
Interestingly, deformations generated by operators of the form given in Eq. \eqref{O2-O1G}, associated with different $G$ functions, mutually commute.
This allows the construction of an infinite commuting family of deformations. 
For example, in two dimensions, the commuting family containing the $T\bar T$ deformation can be generated by the operators 
\begin{equation}
    O_p=t_1 t_2  \left(\frac{t_1-t_2}{t_1 t_2}\right)^p,\quad p\in \mathbb{R}, 
\end{equation}
where $O_0$ is the $T\bar T$ operator and $O_1$ is the root-$T\bar T$ operator.

For a commuting family of stationary deformations, one can choose a basis of homogeneous operators $\{O_p\}$. 
One can show that
\begin{equation}
    \partial_{\tau_p}\big(\sqrt{g}O_q( t_{\alpha,0}) \big) =0,
\end{equation}
for any $p$ and $q$. 
Therefore, the solution of the flow equations is given by
\begin{equation}
    t_\alpha=\frac{t_{\alpha,0}+s}{\sqrt{g/g_0}},\quad  s=\sum_{p}\tau_p(m_p-1) O_p( t_{\alpha,0}),
\end{equation}
where $\sqrt{g}$ can be determined by choosing one of the operators and using its homogeneity property. 
In the holographic context, this solution enables a generalization of the mixed boundary conditions proposal to commuting deformations. 
In the example of the planar AdS black hole, a similar holographic analysis can be performed. The deformed energy satisfies a set of commuting flow equations
\begin{equation}\label{eq:E for commutative}
   \partial_{\tau_p}  E = V O_p(t_\alpha),
\end{equation}
and Eq. \eqref{Tij} remains valid in the presence of commuting deformations, in agreement with the field theory interpretation.

\section{Conclusions}

We proposed a holographic prescription for general stress-energy tensor deformations through mixed boundary conditions. The metric flow equation solutions encode these conditions, as verified for planar AdS black holes. We derived a commuting condition for such deformations and identify infinite commuting families. These results extend previous studies of the holographic $T\overline{T}$ deformation and suggest new routes to engineering holographic dualities.

Several promising directions remain for future work. Extending our analysis beyond the large-$N$ limit is a challenging but important step. One possible approach is to define the deformations by requiring the flow equations of the generating functional to hold exactly, even at finite $N$. 
Recently, the holographic prescription for $T\overline{T}$-deformed conformal field theories with gravitational anomalies was studied in Ref. \cite{Basu:2025fsf}. It would be interesting to apply mixed boundary conditions to more general bulk geometries to include holographic conformal anomalies.

Holography has played a central role in computing many observables in $T\overline{T}$-deformed theories \cite{Tian:2023fgf, He:2023xnb, He:2023hoj, He:2023obo, He:2024pbp, He:2024fdm, Chang:2024voo, Li:2025sfn}. A natural direction is to generalize these calculations to arbitrary stress–energy tensor deformations.
Recent holographic studies have connected $T\bar T$-like deformations to diverse topics, including black hole interiors \cite{AliAhmad:2025kki}, bulk reconstruction \cite{Liang:2025vmx}, and emergent gravity \cite{Adami:2025pqr}. Our results provide a pathway to extend these ideas to general stress-energy tensor deformations in arbitrary dimensions.
Finally, the rich nonperturbative effects of $T\overline{T}$  deformations \cite{Aharony:2018bad,Griguolo:2022xcj, Griguolo:2022hek,Gu:2024ogh, Gu:2025tpy, Hirano:2025alr} raise the question of whether analogous effects occur for general stress-energy tensor deformations and, if so, whether they are related to nonperturbative bulk geometries subject to mixed boundary conditions.

\begin{acknowledgments}

\section{Acknowledgments}
The authors thank Yidian Chen, Song He, Yun-Ze Li, Jia-Rui Sun, Yunfei Xie, Hong-an Zeng, Long Zhao, Zi-Xuan Zhao, and Jie Zhou for useful discussions. 
H. O. is supported by the National Natural Science Foundation of China, Grant No. 12205115 and by the Science and Technology Development Plan Project of Jilin Province of China, Grant No. 20240101326JC.
H. O. wishes to thank the organizers of the “Holographic applications: from Quantum Realms to the Big Bang” conference for their kind hospitality during the completion of this work.
\end{acknowledgments}


\bibliographystyle{simple}
\bibliography{holo2}
\onecolumngrid

\appendix
\section{Supplemental Material}

\subsection{Holographic derivation of flow equations of the energy}
We begin by deriving a useful formula for the solution of the metric flow equation.
We treat $t_\alpha$ as a function of $\tau$ and the initial values $t_{\beta,0}$.
Then we have
\begin{equation}
\begin{split}
  \frac{\partial}{\partial \tau}(\sqrt{g}\frac{\partial}{\partial t_{\beta,0}}t_\alpha) 
  =&  \frac{\partial\sqrt{g}}{\partial \tau}\frac{\partial}{\partial t_{\beta,0}}t_\alpha
  +  \sqrt{g}\frac{\partial}{\partial t_{\beta,0}} \frac{\partial}{\partial \tau}t_\alpha  \\
  =&  \sqrt{g}\sum_{\sigma}\frac{\partial O}{\partial t_{\sigma}}\frac{\partial}{\partial t_{\beta,0}}t_\alpha 
  +  \sqrt{g}\frac{\partial}{\partial t_{\beta,0}} \left( 
  -O+\sum_{\sigma}(t_\sigma-t_\alpha)\frac{\partial O}{\partial t_{\sigma}}
  \right)\\
  =& \sqrt{g} \sum_{\sigma}(t_\sigma-t_\alpha)\frac{\partial}{\partial t_{\beta,0}}\frac{\partial O}{\partial t_{\sigma}}
  \\
  =& \frac{1}{2}\sqrt{g} \sum_{\sigma}(t_\sigma-t_\alpha)\frac{\partial}{\partial t_{\beta,0}}\frac{\partial }{\partial \tau}\log \omega_\sigma
   \\
  =&\frac{\partial }{\partial \tau}\left( \frac{1}{2}\sqrt{g_0} \sum_{\sigma}(t_{\sigma,0}-t_{\alpha,0})\frac{\partial}{\partial t_{\beta,0}}\log \omega_\sigma \right),    
\end{split}
\end{equation}
where we have used $\partial_\tau\left( \sqrt{g}(t_\alpha-t_\beta)\right)  =0$ in the last line. Using the initial conditions $\omega_\alpha=1$ and $t_\alpha=t_{\alpha,0}$ at $\tau=0$, we find
\begin{equation}\label{partial0}
\sqrt{g}\frac{\partial}{\partial t_{\beta,0}}t_\alpha
=\sqrt{g_0}\delta^\beta_\alpha +\frac{1}{2}\sqrt{g_0} \sum_{\sigma}(t_{\sigma,0}-t_{\alpha,0})\frac{\partial}{\partial t_{\beta,0}}\log \omega_\sigma.
\end{equation}
It is straightforward to show that this formula also holds for a commuting family of deformations.

Now we derive the flow equations of the energy of the deformed planar AdS black hole.
For notational simplicity,  it is convenient to perform partial derivative calculations using differentials. The basic variables are $\tau$ and $R^i_{~a}$. 
$h$ and $a_i$ are functions of $\tau$ and $R^i_{~a}$.
$t_\alpha$ and $\omega_\alpha$ are functions of $h$ and $\tau$.
Therefore we have
\begin{equation}\label{dt}
\mathrm{d} \omega_\alpha = (\partial_\tau \omega_\alpha)_h  
\mathrm{d}\tau+\partial_h \omega_\alpha\mathrm{d}h,
~~~
\mathrm{d} t_\alpha = (\partial_\tau t_\alpha)_h  
\mathrm{d}\tau+\partial_h t_\alpha \mathrm{d}h,
\end{equation}
where we denote by $(\partial_\tau...)_h$ the derivative with respect to $\tau$, evaluated while keeping $h$ fixed. 
It follows from the flow equations that
\begin{align}
 (\partial_\tau \omega_\alpha)_h=&2\frac{\partial O}{\partial t_\alpha} \omega_\alpha,\\
  (\partial_\tau t_0)_h=&-O
   +\sum_{\alpha=0}^{d-1} t_\alpha \frac{\partial O}{\partial t_\alpha}
   -t_0 \sum_{\alpha=0}^{d-1}  \frac{\partial O}{\partial t_\alpha}=\frac{d }{\sqrt{g/g_0}}hO_x-O, \\
    (\partial_\tau t_i)_h=&-O
   +\sum_{\alpha=0}^{d-1} t_\alpha \frac{\partial O}{\partial t_\alpha}
   -t_i \sum_{\nu=0}^{d-1}  \frac{\partial O}{\partial t_\alpha}=-\frac{d }{\sqrt{g/g_0}}hO_0-O,
\end{align}
where we denote $O_x=\sum_{i=1}^{d-1}\partial_{t_i}O$ and $O_0=\partial_{t_0}O$.
Using equation (\ref{partial0}) and
\begin{equation}
 \frac{\partial }{\partial h} 
 =(1-d)\frac{\partial }{\partial t_{0,0}}+\sum_{i=1}^{d-1} \frac{\partial }{\partial t_{i,0}},
\end{equation}
we find the relations
\begin{equation}\label{partialh}
\sqrt{g}\partial_h t_0
=\sqrt{g_0}(1-d)+\frac{1}{2}\sum _i \sqrt{g_0}d h \partial_h\log \omega _i,~~~
\sqrt{g}\partial_h t_i
=\sqrt{g_0}-\frac{1}{2}  \sqrt{g_0}d h \partial_h\log \omega _0.
\end{equation}

Now we consider the constraints
\begin{align}
C_0=& V \mathcal{L}^{\frac{d-1}{d}} \Xi\prod_{i=1}^{d-1} {\omega_i^{-1/2}},\\   
C_a=& V a_i\Xi  (t_0-t_x) R^i_{~a}=-\frac{d}{\sqrt{g/g_0}} V a_i\Xi  h R^i_{~a}.
\end{align}
It follows from
\begin{equation}
R^a_{~i}\mathrm{d}\left( \frac{C_a}{C_0} \right)  =0 , 
\end{equation}
that
\begin{equation}
\mathrm{d} a_i
=
a_i \left(\frac{1}{2}\partial_h\log\omega_0
-\frac{1}{d h}\right)
\mathrm{d}h+O_0 a_i \mathrm{d}\tau - R^a_{~i} a_j \mathrm{d} R^j_{~a}.
\end{equation}
Therefore
\begin{equation}\label{dXi}
\mathrm{d}\Xi
=\Xi^{-1} \sum_{i=1}^{d-1} a_i\mathrm{d} a_i
=\frac{(\Xi^2-1) }{\Xi}
\left(
 \left(\frac{1}{2}\partial_h\log\omega_0
-\frac{1}{d h}\right) \mathrm{d}h+O_0   \mathrm{d}\tau
\right)
-\Xi^{-1} R^a_{~i} a^i a_j \mathrm{d} R^j_{~a}.
\end{equation}
It follows from $\mathrm{d} C_0  =0$ that
\begin{equation}
-\frac{1}{2}\sum_{i=1}^{d-1}\partial_h\log\omega_i\mathrm{d}h
-O_x\mathrm{d}\tau+\frac{(d-1) \mathrm{d}h}{d h}+\frac{\mathrm{d}\Xi}{\Xi }+\frac{\mathrm{d}V}{V}=0.\label{dh}
\end{equation}
Using equations (\ref{dt}), (\ref{partialh}) (\ref{dXi}) and (\ref{dh}), we can compute the differential of $E= -V(\Xi^2 t_0-(\Xi^2-1)t_x) $
\begin{equation}
 \begin{split}
\mathrm{d}E=& -\Xi ^2 V \mathrm{d}t_0+\left(\Xi ^2-1\right) V \mathrm{d} t_x + \left(\left(\Xi ^2-1\right) t_x-\Xi ^2 t_0\right)\mathrm{d}V-2 \Xi  V   \left(t_0-t_x\right)  \mathrm{d}\Xi  \\
=&
V O \mathrm{d} \tau  +
 V  \left(-\frac{h d a_i a^j R^a{}_j}{\sqrt{g/g_0}}- t_x R^a{}_i\right)
\mathrm{d}R^i{}_a
\\
=&V O \mathrm{d} \tau  -
 V R^a{}_j T^j{}_i\mathrm{d}R^i{}_a,
 \end{split}   
\end{equation}
which is equivalent to the flow equation of the deformed energy.
For a commuting family of homogeneous deformations, equations (\ref{dt}) become
\begin{equation}
\mathrm{d} \omega_\alpha = 
\sum_p 
(\partial_{\tau_p} \omega_\alpha)_h 
\mathrm{d}\tau_p+\partial_h \omega_\alpha\mathrm{d}h,
~~~
\mathrm{d} t_\alpha =  
\sum_p
(\partial_{\tau_p} t_\alpha)_h  
\mathrm{d}\tau_p+\partial_h t_\alpha \mathrm{d}h,
\end{equation}
with
\begin{equation} 
 (\partial_{\tau_p} \omega_\alpha)_h=2\frac{\partial O_p}{\partial t_\alpha} \omega_\alpha,~~~
  (\partial_{\tau_p} t_0)_h=
  \frac{d }{\sqrt{g/g_0}}hO_{px}-O_p,~~~
    (\partial_{\tau_p} t_i)_h=
    -\frac{d }{\sqrt{g/g_0}}hO_{p0}-O_p.
\end{equation}
Following similar steps, we get
\begin{equation}
  \mathrm{d}E= \sum_p V O_p \mathrm{d} \tau_p  -
 V R^a{}_j T^j{}_i\mathrm{d}R^i{}_a.
\end{equation}

\subsection{Derivation of the commutativity condition}

It follows from
$\partial_{\tau_1}\partial_{\tau_2} \omega_\alpha=\partial_{\tau_2}\partial_{\tau_1} \omega_\alpha$ that
\begin{equation}\label{cm-omega}
\frac{\partial}{\partial \tau_1} \frac{\partial O_2}{\partial t_\alpha}  =\frac{\partial}{\partial \tau_2} \frac{\partial O_1}{\partial t_\alpha}.
\end{equation}
The $\tau_p$-derivatives here account for both explicit and implicit dependencies, while the notation $\tilde\partial_{\tau_p} O_q$ below refers only to the explicit dependence of $O_q$ on $\tau_p$.
Then
$\partial_{\tau_1}\partial_{\tau_2} t_\alpha=\partial_{\tau_2}\partial_{\tau_1} t_\alpha$ leads to
\begin{equation}
\begin{split}
0=& -\partial_{\tau_1} O_2 
+\sum_{\beta}{\frac{\partial}{\partial {\tau_1}}(t_\beta-t_\alpha) \frac{\partial O_2}{\partial t_\beta}}
+\sum_{\beta}{(t_\beta-t_\alpha) \frac{\partial}{\partial {\tau_1}}
\frac{\partial O_2}{\partial t_\beta}}-(1 \leftrightarrow 2)\\
=&-\partial_{\tau_1} O_2 
+\sum_{\beta}{\frac{\partial}{\partial {\tau_1}}(t_\beta-t_\alpha) \frac{\partial O_2}{\partial t_\beta}}-(1 \leftrightarrow 2)\\
=&
-\tilde\partial_{\tau_1} O_2+
\sum_{\sigma}\frac{\partial O_2}{\partial t_\sigma} \left( O_1-
\sum_{\beta}{(t_\beta-t_\sigma) 
\frac{\partial O_1}{\partial t_\beta}}
\right)
-\sum_{\sigma,\beta} {(t_\beta-t_\alpha) \frac{\partial O_1}{\partial t_\sigma}\frac{\partial O_2}{\partial t_\beta}}-(1 \leftrightarrow 2)\\
=&
-\tilde\partial_{\tau_1} O_2+\tilde\partial_{\tau_2} O_1
+
\sum_{\sigma} (O_1 \frac{\partial O_2}{\partial t_\sigma}- O_2 \frac{\partial O_1}{\partial t_\sigma})
  -\sum_{\sigma,\beta} t_\sigma
  \left(
  \frac{\partial O_1}{\partial t_\sigma}
  \frac{\partial O_2}{\partial t_\beta}
  -\frac{\partial O_2}{\partial t_\sigma}
  \frac{\partial O_1}{\partial t_\beta}
  \right)\\
  =&C(O_1,O_2),\label{cm-ti}
\end{split}
\end{equation}
where we have used (\ref{cm-omega}) in the second line.
Taking the derivative of $C(O_1,O_2)=0$ with $t_\alpha$, we obtain (\ref{cm-omega}). Therefore, two deformations commute with each other if and only if $C(O_1,O_2)=0$.

\end{document}